\documentclass{PoS}
\def\gsim{\:\raisebox{-0.5ex}{$\stackrel{\textstyle>}{\sim}$}\:}
\def\lsim{\:\raisebox{-0.5ex}{$\stackrel{\textstyle<}{\sim}$}\:}

\title{Dark Matter Theory}

\ShortTitle{Dark Matter Theory}

\author{\speaker{Manuel Drees}\\
  Physikalisches Inst. and BCTP, Bonn University, Nussallee 12, 53155
  Bonn, Germany\\
        E-mail: \email{drees@th.physik.uni-bonn.de}}

      \abstract{I begin by briefly reviewing the evidence for the
        existence of Dark Matter (DM), emphasizing that {\em many}
        observations, at length scales between kpc (the size of the
        smallest galaxies) and Gpc (the Hubble radius) can be
        described by the same simple model, $\Lambda$CDM.  I will then
        argue that primordial black holes, the only DM candidates that
        can be realized within the Standard Model (SM) of particle
        physics, are very unlikely to provide all of DM. After giving
        a (probably incomplete) list of possible DM candidates, I end
        by mentioning some recent developments in the theory of Weakly
        Interacting Massive Particles (WIMPs).}

\FullConference{The 39th International Conference on High Energy Physics (ICHEP2018)\\
		4-11 July, 2018\\
		Seoul, Korea}

\begin{document}

\section{Evidence for Dark Matter}

For a cosmologist, ``matter'' is stuff whose pressure $p$ is much less
than its energy density $\rho$ (in natural units where
$\hbar = c = 1$). In contrast, ``radiation'' is stuff with
$p \simeq \rho/3$, and ``dark energy'' is stuff with negative
pressure, $p < -\rho/3$ (the simplest example is vacuum energy with
$p = -\rho$, which is indistinguishable from Einstein's cosmological
constant). In the matter category, we need to further distinguish
between ``baryonic'' matter, which consists of protons, neutrons and
electrons (yes, electrons are baryons for cosmological purposes -- but
neutrinos aren't), and Dark Matter, which doesn't. (This is the modern
definition of DM. It differs somewhat from the older definition as
``matter which cannot be detected optically''. Most, but not all, DM
by the old definition is also DM by the new definition, but by the old
definition there's also baryonic DM; more on that below.) If we assume
that gravitational interactions at all length scales $\gsim 1$ kpc
(about $54$ orders of magnitude above the Planck length, so most
likely we need not worry about quantum gravity) is described by
Einstein's general theory of relativity, a great many {\em
  independent} observations require the existence of DM \cite{pdgrev}:
\begin{itemize}
\item {\bf Galactic rotation curves:} Assuming (approximately)
  spherical symmetry, bodies on stable Keplerian orbits with radius
  $r$ around a galaxy have rotational (tangential) velocity
  $v(r) \propto M(r)/r$, where $M(r)$ is the mass within the orbit.
  Observationally, $v(r)$ remains rather flat (although not exactly
  constant) even at radius $r$ well outside of nearly the entire
  visible mass of the galaxy. This can be explained by the presence of
  a dark component, such that $M(r) \propto r$ (approximately).
\item {\bf Clusters of galaxies:} The mass of a cluster of galaxies
  can be estimated in different ways: via the relative velocities of
  the galaxies in the cluster (assuming the system is virialized); via
  the temperature of the hot, X--ray emitting gas that contains most
  of the baryons in the cluster\footnote{Zwicky used the first method
    to estimate the mass of the Coma cluster. Back in the 1930's
    X--ray emitting matter was invisible to astronomers, so this hot
    gas was counted as part of Zwicky's DM. This illustrates a problem
    with baryonic DM: the observational definition of ``dark'' depends
    on technology, and is hence time--dependent; the distinction
    between baryonic and non--baryonic matter isn't.}; and, most directly,
  via gravitational lensing of background galaxies. In all cases one finds
  that the total mass of the cluster exceeds its baryonic mass by about a
  factor of $5$.
\item{\bf Old galaxies:} We know from the cosmic microwave background
  (CMB) that the universe was very homogeneous at temperature
  $T \simeq 0.3$ eV. Small density perturbations can grow (by
  overdensities gravitationally attracting more matter) only in a
  matter--dominated epoch. Large overdensities, as required for galaxy
  formation, can therefore occur earlier for larger matter content. In
  particular, given the size of fluctuations in the CMB, the observations
  of galaxies at $z \gsim 10$ can only be understood if a sizable amount
  of DM exists.
\item{\bf Bullet cluster:} This is actually a system of two clusters,
  which ``recently'' passed through each other with rather large relative
  velocity. The hot plasma in the two clusters interacted strongly,
  heating up and slowing down. As a result, the plasma clouds are now
  well separated from the bulk of the galaxies in the cluster. At the same
  time gravitational lensing shows that the total mass of the clusters
  centers at the same location as the galaxies do -- not where most baryonic
  matter, stored in the hot X--ray emitting plasma, is located. In the context
  of DM, this can be used to derive an upper bound on the DM--DM scattering
  cross section \cite{bullet},
  \begin{equation} \label{e1}
    \sigma_{\chi \chi} \leq ({\rm few}) \ {\rm b} \cdot \frac {m_\chi}
    {1 \ {\rm GeV}}\,,
  \end{equation}
  where $\chi$ stands for a DM particle.\footnote{Soon after the
    discovery of this system it was claimed \cite{claim} that it
    argues against the standard $\Lambda$CDM cosmology, because
    simulations did not produce encounters of similarly large clusters
    with similarly high relative velocity. Such systems are indeed
    rare in the observable universe: no comparable system has yet been
    found. It seems that simulations available at that time did not
    encompass a sufficiently large volume to include similar
    systems. More recent simulations conclude \cite{counterclaim} that
    $\Lambda$CDM has no problem generating such a system.} Note that the
  $pp$ and $p \bar p$ scattering cross sections are well below this
  bound; pions come closer to saturating it.
\item{Acoustic oscillations:} In addition to fixing the normalization
  of density perturbations at $T \simeq 0.3$ eV, the analysis of CMB
  anisotropies also allows to measure the ``sound horizon'', showing
  that the visible universe is essentially spatially flat, and to
  determine the free parameters of the $\Lambda$CDM model with
  percent--level precision. In particular, the overall relic density
  of DM is given by \cite{planck18}
  \begin{equation} \label{e2}
    \Omega_{\chi} h^2 = 0.1198 \pm 0.0012\,.
  \end{equation}
  Here $\Omega_\chi$ is the $\chi$ mass density in units of the
  critical density, and $h\simeq 0.7$ is today's Hubble constant in
  units of $100$ km$/$(s$\cdot$Mpc). It should be noted that the most
  prominent of these acoustic oscillations have also been detected in
  galaxy correlation functions at low redshift.
\end{itemize}

While $\Lambda$CDM can account for the large--scale structure of the
visible Universe very well, there are some challenges at smaller
scales \cite{challenges}, often involving dwarf galaxies. All these
problems occur in systems with very high density contrast (so that
density fluctuations can no longer be treated perturbatively) {\em
  and} where baryonic physics (sometimes called ``baryochemistry'' by
cosmologists) is likely to play an important role. The latter is
currently not sufficiently well understood to decide whether all
observations can be reproduced by the $\Lambda$CDM model.

However, the ``Lyman $\alpha$ forest'' -- the spatial distribution of
hydrogen clouds, detected through the absorption of photons at the
Ly$\alpha$ wave length -- {\em is} very well described by $\Lambda$CDM
\cite{challenges}; these are much simpler systems, at similar length
scales as small galaxies. This success strongly constrains extensions
of the $\Lambda$CDM model (e.g., interacting DM) that is meant to
improve the description of small galaxies. Therefore currently there
is no compelling need to go beyond $\Lambda$CDM.

So far we have assumed that GR describes gravitational interactions at
the relevant length scales. Many of the observations alluded to above
(e.g., of galactic rotations curves and peculiar velocities in
clusters) primarily indicate a missing force, i.e. the force on the
observed objects is larger than predicted by GR if all the mass is in
the visible baryons. Within the framework of GR this can be fixed by
adding mass, i.e. DM. Alternatively, one can contemplate changing the
law of gravity. In fact, MOdified Newtonian Dynamics (MOND)
\cite{mond} as a purely phenomenological modification of Newton's law
of gravity at small accelerations can account quite well for galactic
rotation curves.

However, embedding this into a consistent covariant (classical) field
theory is quite difficult: it is not easy to devise an interaction
whose potential falls off more slowly than $1/r$ at large distances,
as required if MOND is to explain the observations. Existing attempts
include either an additional vector and an additional scalar field
(``TeVeS'') \cite{teves}, or a second tensor (``bi--metric gravity'')
\cite{bimetric}. Moreover, observations of the bullet cluster, and
cosmological observations, seem to require some sort of DM even in
MOND \cite{mond_dm}. Hence, MOND is certainly no more economical than
GR plus DM; besides, GR can explain a large body of data, including
some of the most precise measurements ever performed (on binary
neutron stars) \cite{grtest} and the recent observation of
gravitational waves \cite{ligo}. In contrast to these theoretical
difficulties with MOND, DM can be introduced using the
well--established formalism of quantum field theory; it thus seems the
far more plausible explanation to me.

\section{Properties of Dark Matter}

Many of the observations requiring the existence of DM are made in the
nearby universe. This implies that DM particles must be extremely
long--lived, $\tau_\chi \gsim 10^{10}$ years. For DM particles heavier
than a few GeV the bound on the lifetime is even stronger, typically
$\tau_\chi \gsim 10^{25}$ seconds \cite{decay}; this comes from unsuccessful
searches for the energetic $\chi$ decay products. Most (but not all)
DM models therefore assume $\chi$ to be absolutely stable.

By definition DM should be ``dark'', meaning it should not interact
(much) with photons. The simplest choice is to consider neutral
particles, of course, but ``millicharged'' particles with electric
charge $ |q_\chi| \lsim 10^{-3} e$ may also be acceptable
\cite{milli}. The strongest bound on charged stable particles comes
from the search for exotic isotopes \cite{stable}. A massive particle
with charge $-e$ could bind to some nucleus ${}^Z_AX$. For
$m_\chi > 10$ GeV or so the binding energy would exceed $10$ MeV even
for quite light nuclei (e.g. carbon or oxygen). The bound state would
thus chemically look like an isotope with charge $Z-1$ but with mass
$\simeq A m_p + m_\chi$. Large--scale searches for such exotic
isotopes have failed to find anything. For example at most one out of
$5 \cdot 10^{15}$ carbon nuclei can be bound to a charged particle of
mass $10$ TeV. For yet heavier particles, the best limit comes from
looking for superheavy water, made from hydrogen--like $\chi^+ e^-$
``atoms''. It seems plausible that similar limits apply to stable
massive particles that carry color by no electric charge, since they
should bind to ordinary nuclei as well.

Of course, in order to count as matter, $\chi$ must have
non--vanishing mass. Moreover, simulations of galaxy formation
strongly favor ``cold'' dark matter \cite{challenges}, i.e. $\chi$
must have been non--relativistic at the latest when one Hubble volume
contained a galactic mass. For particles that have been produced more
or less thermally, this implies a lower bound on the mass in the keV
range. Similarly, fermionic DM, which is subject to the Pauli
exclusion principle, must have a mass at least in the keV range in
order to be able to accommodate a sufficient amount of mass in the
limited phase space of a dwarf galaxy \cite{tg}. On the other hand,
bosonic DM that never was in thermal equilibrium might be much lighter
than this. The lower bound than comes from the requirement that the
de~Broglie wavelength $1/(v_\chi m_\chi)$ (with $v_\chi \lsim 10^{-3}$
in small galaxies) should not be larger than the scale length of the
smallest ``known'' DM halos, those of dwarf galaxies \cite{fuzzy};
this gives a lower bound of $10^{-23}$ eV or so. To summarize:
\clearpage
\begin{eqnarray} \label{e3}
  10^{-23}\ {\rm eV} & \lsim m_\chi & \lsim M_{\rm Pl} \
                           ({\rm non-thermal, \ bosonic \ DM}); \nonumber \\
       1 \ {\rm keV} & \lsim m_\chi & \lsim M_{\rm Pl} \
                                      ({\rm thermal \ or \ fermionic \ DM}).
\end{eqnarray}
Here $M_{\rm Pl} \simeq 2.4 \cdot 10^{18}$ GeV is the reduced Planck
mass; I used it as (approximate) upper bound since I find it hard to
envision elementary particles with even larger mass. Even if we accept
this as upper bound, the range (\ref{e3}) spans $50$ orders of
magnitude for DM bosons that were not produced thermally, and about
$24$ orders of magnitude for other candidates.
                                      
Finally, we have the (quite weak) upper bound (\ref{e1}) on the
$\chi \chi \rightarrow \chi \chi$ scattering cross section, which is
actually not easy to violate in concrete models. Of course, the DM
density is by now known quite accurately, at least within
$\Lambda$CDM, see eq.(\ref{e2}). (In simple extensions of $\Lambda$CDM
the errors become a bit larger, but the central value doesn't change
much.) However, in general this does not translate directly into a bound
on a property, or a combination of properties, of $\chi$.

We are thus forced to conclude that, while we have {\em very solid
  evidence} for the existence of DM, we {\em know very little} about
its properties.  Remarkably enough, the above list of known properties
suffices to exclude all elementary particles described by the SM as
primary component of DM. In particular, the only known massive,
non--baryonic particles -- the (active) neutrinos -- are way too
light.

There is one possible loophole in this argument. The earliest
determination of the baryon content of the Universe comes from
analyses of Big Bang Nucleosynthesis (BBN). While less accurate than
the constraint from the CMB data, it also implies that the baryonic
density cannot be much more than $5\%$ of critical density, which is
well below what is needed for the DM density. However, baryons that
were bound in black holes at the time of BBN do not count as baryons
any more. Such ``primordial'' black holes (PBHs) \cite{carr} even have
some independent motivation: as early seeds of the superheavy black
holes seen in AGNs (formerly known as quasars) already at quite high
redshift; and as explanation of the fact that all black holes mergers
that have been observed via their gravitational wave signature so far
have quite large masses, perhaps larger than what prior modeling of
``population III'' stars indicated. (Somewhat counter--intuitively,
``population I'' refers to the youngest stars. ``Population III''
stars would have had to form from primordial matter, basically only
consisting of hydrogen and helium, i.e.  with essentially vanishing
``metallicity'' in astronomers' language.  Such stars are thought to
have been very massive, and hence short--lived; this explains why no
such stars can be found in our ``neighborhood''.) This triggered quite
a lot of work on constraining PBHs as possible DM candidates using a
multitude of astronomical observations and astrophysical
considerations. By now it seems clear that only PBHs with mass near
$10^{16}$ to $10^{17}$ g can form a sizable fraction of DM
\cite{pbhrange}. The density of PBHs with slightly lower mass is very
strongly constrained because such objects would evaporate via Hawking
radiation just about now, which would give very spectacular
signals. PBHs can form only if the primordial density perturbations at
the relevant (very small) length scales are at least $7$ orders of
magnitude larger than those probed by the CMB; the deviation of scale
invariance observed in the CMB goes in the {\em wrong} direction. This
raises two problems: one has to find a mechanism that generates very
strong primordial density perturbations at length scales corresponding
to $10^{17}$ g PBHs; and the density perturbations at slightly smaller
length scales, which would produce PBHs decaying now, has to be
considerably smaller again.  While this is technically possible, it
seems very contrived. Very recently it has also been pointed out that
models that are engineered accordingly will generate a detectable
amount of gravitational waves in the LISA band \cite{pbh_gw}. In
summary, severe finetuning is required for PBHs to form a significant
fraction of DM; and models of this kind can soon be tested.

\section{Particle Physics Candidates}

Given how little we know about the properties of DM (see the previous
Section) it is not surprising that a large number of different
particle physics candidates have been suggested. These cover the
entire allowed range (\ref{e3}) of masses, from ``fuzzy'' \cite{fuzzy}
to ``Planckian'' \cite{planckian} DM; their self interactions range
from essentially negligible (in the large majority of models) to
saturating the bound (\ref{e1}). However, I personally still prefer
candidates that have some {\em independent} motivation, not related to
DM, or even to cosmology. This is true in particular for axions
\cite{ax_rev}, which have been introduced so solve the strong CP
problem but can also make good CDM if the ``Peccei--Quinn'' scale
where the global $U(1)$ symmetry of the axions is broken is around
$10^{10}$ GeV. It is also true for some Weakly Interacting Massive
Particles (WIMPs) \cite{pdgrev} that emerge in attempts to solve, or
at least alleviate, the hierarchy problem of the Standard Model, the
prime (but not quite only) example here being supersymmetric WIMPs
\cite{susy_rev}. To the best of my knowledge axions are still
perfectly viable CDM candidates; but since I have never worked on
them, for the rest of this write--up I will focus on WIMPs.

\section{News on WIMPs}

In the last couple of decades, WIMPs have been the by far most popular
DM candidates, to the extent that sometimes WIMP and DM are taken to be
synonyms. They aren't -- as I just mentioned, axions are perfectly fine
CDM candidates, and they are certainly not WIMPs -- but the popularity
of WIMPs is not without reason.

In fact, there are two reasons, a more theoretical and a
phenomenological one. The theoretical argument in favor of WIMPs rests
on the fact that the simplest production mechanism for massive relics
from the very early universe, freeze--out from the thermal plasma
\cite{kt}, in minimal cosmology automatically points towards the weak
scale. The main assumption here is that the post--inflationary
universe once was sufficiently hot and dense that reactions of the kind
$\chi \chi \rightarrow {\rm SM}$ were in equilibrium. Note that in a
static universe {\em all} reactions eventually reach equilibrium.
However, in an expanding universe equilibrium can be maintained only
if the reaction rate, given by
$n_\chi \langle \sigma_{\chi\chi} v \rangle$, is larger than the
expansion rate, given by the Hubble parameter,
$H \propto T^2 / M_{\rm Pl}$ in the radiation--dominated era.  Here
$n_\chi$ is the $\chi$ number density, $\sigma_{\chi\chi}$ is the
total cross section for the annihilation of two DM particles into an
SM final state, $v$ is the relative velocity between the
annihilating DM particles, and $M_{\rm Pl}$ is the Planck mass.  Due
to the $1/M_{\rm Pl}$ factor in $H$, if the couplings of $\chi$ are
not very tiny the reaction rate will be larger than the expansion rate
for temperature $T \sim m_\chi$. However, if $\chi$ is in equilibrium,
$n_\chi \propto {\rm e}^{-m_\chi / T}$ for $T < m_\chi$ is exponentially
suppressed, and will thus become smaller than $H$ eventually; for typical
WIMPs this ``freeze--out'' occurs at $T \sim m_\chi / 20$ (with logarithmic
dependence on $\langle \sigma_{\chi\chi} v \rangle$). After freeze--out
the number of $\chi$ particles per co--moving volume is basically constant.
This gives the final result
\begin{equation} \label{thermal}
\Omega_\chi h^2 \propto \frac {1} {\langle \sigma_{\chi\chi} v \rangle}\,.
\end{equation}  
The observational value (\ref{e2}) is saturated for weak--scale (i.e.,
picobarnish) cross section. This is sometimes\footnote{Pompously or
  derisively, I'm not sure.} called the ``WIMP miracle''. While there
is nothing miraculous (meaning super--natural) about this result, it
is at the very least a coincidence linking weak--scale physics with
DM.\footnote{This argument assumes that the main $\chi$ number changing
  reactions are $\chi\chi$ annihilations into SM particles. In some models
  the main number changing reactions are of the form $3\chi \leftrightarrow
  2\chi$, i.e. occur purely within the dark sector. The desired relic density
  (\ref{e2}) is then obtained for $\chi$ mass and interaction strength
very roughly comparable to those of ordinary pions, with $2 \rightarrow 2$
scattering cross sections not far below the bound (\ref{e1}) \cite{simp}.}
The fact that naturalness arguments favor the existence of new weak--scale
particles makes this coincidence all the more intriguing. This is the
theoretical argument in favor of WIMPs.

WIMPs are interesting phenomenologically because a roughly weak--scale
annihilation cross section indicates that both the annihilation of
WIMPs in the current universe and the elastic scattering of WIMPs on
target nuclei might be detectable; the former goes under indirect
detection, while the latter constitutes direct detection
\cite{pdgrev}. In today's universe WIMP annihilation can only be
detectable in regions with enhanced WIMP density, e.g. near the
centers of galaxies or within the Sun (which can capture WIMPs that
have lost energy by elastically scattering off a nucleus in the
Sun). However, this argument is not airtight. For example, if at low
$v$ $\chi \chi$ annihilation from an $S-$wave initial state is
suppressed, then
$\langle \sigma_{\rm \chi\chi} v \rangle \propto v^{2n}$ with
$n \geq 1$ ($n=1$ for annihilation from a $P-$wave), and the average
$v^2$ in our galaxy is about 5 orders of magnitude smaller than that
during WIMP decoupling. Moreover, the relic density might actually be
determined by co--annihilation of $\chi$ with another slightly heavier
partner particle (e.g. a squark if $\chi$ is the LSP in the MSSM)
\cite{gs}; by now these partner particles will all have decayed, and
the true $\chi\chi$ annihilation cross section can be very small in
such a scenario.  On the other hand, direct WIMP detection can also be
suppressed, e.g.  if $\chi$ predominantly couples to top quarks, in
which case $\chi$ couples to nucleons only at one--loop level.

Nevertheless the fact that even the recent ton--scale direct detection
experiments, e.g. Xenon-1t \cite{x1t}, have not found a signal puts
severe strain on many well--motivated WIMP candidates. This has led to
a lot of activity in model--building, which in turn is starting to spur
new experimental developments.

There are basically three ways to evade the current direct detection
bounds. One is to go for very light WIMPs. They cannot be detected
with current methods since a WIMP that is much lighter than the target
nucleus cannot transfer a detectable amount of energy to it (just like
a table tennis ball hitting a basket ball will just bounce back, leaving
the basket ball at rest). Such models typically need new ``mediator''
particles to facility WIMP annihilation into SM particles with a sufficiently
large cross section. Both the light DM particles (which at some point
become too light to be properly called WIMPs) and the mediators are
being searched for in a large variety of (relatively low--energy)
experiments.

The second option is to go for very heavy WIMPs. Here the direct
detection bound on the scattering cross section weakens because the
bound is really on the event rate, which is proportional to the
product of WIMP flux and scattering cross section, and for given WIMP
mass density in the solar neighborhood the WIMP flux drops
$\propto 1/m_\chi$. Hence for $m_\chi > 100$ GeV or so the bound on
the scattering cross section scales basically linearly with
$m_\chi$. Of course, for dimensional reasons the annihilation cross
section will scale like $1/m_\chi^2$. For very large $m_\chi$ it thus
becomes difficult to design the model such that the $\chi$
annihilation cross section is sufficiently large (in minimal
cosmology). In fact, unitarity arguments \cite{gk} imply the upper
limit $m_\chi \leq 120$ TeV. The largest mass of a thermal WIMP in
minimal cosmology that has been found \cite{df} in a reasonably well
motivated particle physics model (the MSSM extended with right--handed
sneutrino superfields an an extra $U(1)$ gauge factor) is about $40$
TeV, which is already within a factor of $3$ of the unitarity bound.

The final option is to consider WIMPs with very small scattering cross
section on nuclei. This is e.g. true for Bino--like neutralinos (i.e.
superpartners of the hypercharge gauge boson) in the MSSM. These are
gauge singlets, which couple to Higgs bosons only via mixing with
higgsinos; this mixing is small if the higgsinos are heavy. Binos can
scatter also via squark exchange, but LHC experiments tell us that
squark masses are in the TeV range at least \cite{squark}, so these
contributions are also small.

However, because Binos have such small couplings, their relic density
in minimal cosmology typically comes out (much) too large. One way
to fix this is to consider co--annihilation with colored superparticles,
the most widely candidate being the lighter stop squark \cite{bdd}.
Recent calculations show that this can give the correct relic density
for Binos with mass up to $6$ TeV \cite{bl}. However, co--annihilation
requires that the stop and the Bino have similar masses, the mass difference
being below $10\%$; Bino masses near $6$ TeV even require a mass splitting
$\lsim 0.5\%$, which appears rather finetuned.

Another possibility is to consider deviations from minimal cosmology.
In particular, if for some range of temperatures above MeV (i.e.,
safely above the range relevant for Big Bang nucleosynthesis) the
energy density of the universe was dominated by some heavy but long--lived
field many possibilities open up \cite{gg}. ``Moduli'' fields with
these properties in fact occur quite naturally in many string theory
constructions \cite{kane}. Detailed calculations show that in this
case Binos can indeed be good DM candidates in generic parts of
MSSM parameter space if the branching ratio for moduli decays into
superparticles is very small, below $10^{-4}$ or so \cite{dh}.

\section{Take--Home Messages}

\begin{itemize}

\item Dark Matter exists!

\item Most likely it consists of new kind(s) of particle.

\item We don't know much about these particles.

\item Several decades of searches for Dark Matter have not yielded
  a convincing signal.

\item Some of the oldest/simplest/best motivated candidates (axion,
  WIMPs, gravitinos) are {\em not} excluded.

\item Since WIMPs are getting squeezed, theorists are extending WIMP
  parameter space, and are suggesting entirely new kinds of DM
  candidates, with masses between $10^{-22}$ eV and $10^{18}$ GeV.

\end{itemize}

\vspace*{3mm} \noindent
{\bf Acknowledgments:} I thank the organizers for the opportunity to
give a plenary talk at the place where I got married more than 18
years ago. I also wish to apologize to everybody whose highly relevant
work could unfortunately not be cited due to limitations of space.


\begin{thebibliography}{99}
\bibitem{pdgrev}
See e.g. the Review on Dark Matter in the Particle Data Book,
M. Tanabashi et al. (Particle Data Group), Phys. Rev. {\bf D 98} (2018)
030001 (2018).
%
\bibitem{bullet}
A. Robertson, R. Massey and V. Eke, MNRAS {\bf 465} (2017) 569
[arXiv:1605.04307].
%
\bibitem{claim}
J. Lee and E. Komatsu, ApJ {\bf 718} (2010) 60 [arXiv:1003.0939].
%
\bibitem{counterclaim}
C. Lage and G.R. Farrar, JCAP {\bf 02} (2015) 38 [arXiv:1406.6703];
R. Thompson, R. Dav\'e and K. Nagamine, MNRAS {\bf 452} (2015)
3030 [arXiv:1410.7438].
%
\bibitem{planck18}
Planck Collab., N. Aghanim et al., arXiv:1807.06209.
%
\bibitem{challenges}
For a review, see e.g. M.R. Buckley and A.H.G. Peter, Phys. Rep. {\bf 761}
(2018) 60 [arXiv:1712.06615].
%
\bibitem{mond}
M. Milgrom, Can. J. Phys. {\bf 93} (2015) 107-118 [arXiv:1404.7661].
%
\bibitem{teves}
J.D. Bekenstein, Phys. Rev. {\bf D70} (2004) 083509 [arXiv:astro-ph/0403694].
%
\bibitem{bimetric}
M. Milgrom, Phys. Rev. {\bf D80} (2009) 123536 [arXiv:0912.0790].
%
\bibitem{mond_dm}
  L. Bernard and L. Blanchet, Phys. Rev. {\bf D91} (2015) 103536
  [arXiv:1410.7708].
%
\bibitem{grtest}
E. Berti et al., Class. Quant. Grav. {\bf 32} (2015) 243001
[arXiv:1501.07274].
%
\bibitem{ligo}
B.P.Abbott et al., LIGO and Virgo Collabs., Phys. Rev. Lett. {\bf 116} (2016)
061102 [arXiv:1602.03837].
%
\bibitem{decay}
J.R. Ellis, G.B. Gelmini, J.L. Lopez, D.V. Nanopoulos and S. Sarkar,
Nucl.Phys. {\bf B373} (1992) 399.
%
\bibitem{milli}
E. Del Nobile, M. Nardecchia and P. Panci, JCAP {\bf 1604} (2016) 048
[arXiv:1512.05353].
%
\bibitem{stable}
See the review on Other Particle Searches in the Particle Data Book [1].
%
\bibitem{tg}
S. Tremaine and J.E. Gunn, Phys. Rev. Lett. {\bf 42} (1979) 407;
C. Di Paolo, F. Nesti and F.L. Villante, MNRAS {\bf 475} (2018) 5385
[arXiv:1704.06644].
%
\bibitem{fuzzy}
A. Arbey, J. Lesgourgues and P. Salati, Phys. Rev. {\bf D64} (2001) 123528
[arXiv:astro-ph/0105564].
%
\bibitem{carr}
B.J. Carr and S.W. Hawking, MNRAS {\bf 168} (1974) 399.
%
\bibitem{pbhrange}
B. Carr, M. Raidal, T. Tenkanen, V. Vaskonen and H. Veerm{\"a}e,
Phys. Rev. {\bf D96} (2017) 023514 [arXiv:1705.05567].
%
\bibitem{pbh_gw}
N. Bartolo et al., arXiv:1810.12224.
%
\bibitem{planckian}
K.A. Meissner and H. Nicolai, arXiv:1809.01441.
%
\bibitem{ax_rev}
See e.g. the review on Axions in the Particle Data Book [1].
%
\bibitem{susy_rev}
See e.g. the review on Supersymmetry in the Particle Data Book [1].
%
\bibitem{kt}
E.W. Kolb and M.S. Turner, {\it The Early Universe}, Front. Phys. 69 (1990).
%
\bibitem{simp}
See e.g. Y. Hochberg, E. Kuflik, H. Murayama, T. Volansky and J.G. Wacker,
Phys. Rev. Lett. {\bf 115} (2015) 021301 [arXiv:1411.3727];
N. Bernal, X. Chu and J. Pradler, Phys. Rev. {\bf D95} (2017) 115023
[arXiv:1702.04906];
S.-M. Choi et al., JHEP {\bf 1710} (2017) 162 [arXiv:1707.01434].
%
\bibitem{gs}
K. Griest and D. Seckel, Phys. Rev. {\bf D43} (1991) 3191.
%
\bibitem{x1t}
E. Aprile et al., XENON Collab., Phys. Rev. Lett. {\bf 121} (2018) 111302
[arXiv:1805.12562].
%
\bibitem{gk}
K. Griest and M. Kamionkowski, Phys. Rev. Lett. {\bf 64} (1990) 615.
%
\bibitem{df}
M. Drees and F.A. Gomes Ferreira, to appear.
%
\bibitem{squark}
See e.g. ``Supersymmetric Particle Searches'' in the Particle Data book [1].
%
\bibitem{bdd}
C. Boehm, A. Djouadi and M. Drees, Phys. Rev. {\bf D62} (2000) 035012,
[hep-ph/9911496].
%
\bibitem{bl}
S. Biondini and M. Laine, JHEP {\bf 1804} (2018) 072 [arXiv:1801.05821].
%
\bibitem{gg}
G.B. Gelmini and P. Gondolo, Phys. Rev. {\bf D74} (2006) 023510
[hep-ph/0602230].
%
\bibitem{kane}
B.S. Acharya, G. Kane, S. Watson and P. Kumar, Phys. Rev. {\bf D80} (2009)
083529 [arXiv:0908.2430].
%
\bibitem{dh}
M. Drees and F. Hajkarim, JCAP {\bf 1802} (2018) 057 [arXiv:1711.05007],
and arXiv:1808.05706.
%




\end{thebibliography}
\end{document}